\documentclass{article}       
\usepackage{e}                
\input psfig
\begin{document}
\branch{B}   
%
\title{The Labusch Parameter of a Driven 
Flux Line Lattice in YBa$_2$Cu$_3$O$_7$
Superconducting Films}
\author{Alexey V. Pan and Pablo Esquinazi}
\institute{Department of  Superconductivity and Magnetism,\\ Institut f\"ur Experimentelle Physik II, Universit\"at  Leipzig,\\
Linn{\'e}strasse 5, D-04103 Leipzig, Germany\\ e-mail: esquin@physik.uni-leipzig.de}
\titlerunning{The Labusch Parameter of a Driven 
Flux Line Lattice}
\maketitle
\begin{abstract}
We have investigated 
the influence
of a driving force on the elastic coupling (Labusch parameter) 
of the field-cooled state of
the flux line lattice (FLL) in 400 nm thick YBa$_2$Cu$_3$O$_7$
superconducting films.
We found that the FLL of a field-cooled state without driving forces 
is not in an equilibrium state. Results obtained for magnetic fields
applied at $0^\circ$ and 30$^\circ$ relative to CuO$_2$ planes, show
an enhancement of the elastic coupling of the films
at driving current densities several orders of magnitude smaller
than the critical one. Our results indicate that the  FLL 
appears to be in a relatively ordered, metastable state after
field cooling without driving forces.
\end{abstract}

\section{Introduction}
The investigation of effects directly related to true phase trasitions of
the flux line lattice (FLL) in high-temperature superconductors (HTS) is
 a current topic of research. These studies are not only
difficult to perform experimentally due to the necessary resolution but also 
their interpretation is sometimes not straightforward due to 
pinning effects. In the quasi three dimensional (3D) HTS YBa$_2$Cu$_3$O$_7$ 
(Y123) an anomalous enhancement of the pinning force appears to be
associated with the ``peak
effect" (PE) in the dc critical current measured using the voltage criterium
 \cite{kowk}. In contrast to conventional low-$T_c$
superconductors, this PE occurs at fields much below the
upper critical field, near the thermally activated depinning line 
\cite{zieseprb}.
Within the classical argument of Pippard \cite{pipa} 
the PE in a 3D superconductor with a high density 
of pinning centers 
occurs if at high enough fields or temperatures 
 the shear modulus or rigidity of the FLL strongly drops. This softening of the FLL 
observed increasing temperature and/or
magnetic field should overwhelm the usual decrease of the pinning
force in order to have an effective enhancement of the pinning
force. Within this picture the correlation between the PE and a
reduction of the rigidity of the FLL is appealing.

The experimental evidence for a correlation between the PE
and the ``melting" of the FLL is, however,  far from 
being clear. As noted in Ref.~\cite{hb} the sharp {\em enhancement} of resistance
with temperature in Y123 crystals, interpreted as a first-order 
melting transition \cite{safar},  is precisely opposed to the
enhancement of pinning expected within Pippard's picture. The results of
a remarkable experimental study characterising the
 PE in a  clean and anisotropic superconductor suggested that
much of the transport data interpreted as melting in HTS may
be related to nonequilibrium dynamical steady states of the
FLL and not to true phase transitions \cite{hb}.  

Studies of the PE and its dependence on the magnetic field history in Y123
thin films and crystals using the vibrating reed technique \cite{zieseprb}
casted some doubts on the interpretation of the PE line as a melting
line. It was suggested that the onset of the PE is due to the thermal
creation of FLL dislocations and that some of the melting lines may
be observations of the PE using different criteria \cite{zieseprb}.
The strongly nonlinear response of  vibrating superconductors near
the PE \cite{zieseprb} is qualitatively consistent with the dynamic phase
diagram measured in \cite{hb}. 

New features  recently found in Y123 crystals by transport
experiments reveal a PE at high fields in the vortex
solid region of the phase diagram \cite{crab}. This PE
joins smoothly the -- believed to be -- first-order melting line at high temperatures
\cite{crab} and suggests a correlation with the previously observed
PE at low fields \cite{kowk,zieseprb}. The rather temperature
independence of this new PE at high fields was speculated to be
the analog of the second magnetization peak observed
in Bi$_2$Sr$_2$CaCu$_2$O$_8$ (Bi2212) HTS \cite{crab}. Recently published
measurements on Y123 crystals up to 27~T reveal that the temperature
dependence of the field at the
second magnetization peak is qualitatively similar to that observed in 
 Bi2212 crystals \cite{japansmp}. Depending on the crystal
properties, the second magnetization peak can be observed even in
the same temperature range,  i.e. at $T < 50~$K, as in Bi2212 HTS \cite{nisi2}.
Recent experimental
studies cast  strong doubts about the interpretation of the
second magnetization peak in Bi2212 HTS based on an
pinning enhancement and related to a phase transition of 
the FLL \cite{kope1,kope2}. The origin of the so-called
``pre-melting" peaks in the critical current density obtained
by transport measurements is therefore still unclear \cite{crab}. 

As the vibrating reed results in Y123 crystals and thin films
showed, the observed PE depends on the magnetic history of the sample
as well as on the driving force \cite{zieseprb} provided by the reed amplitude, which
in general is $\le 150~$nm. The observed history effect means that the magnitude of 
the PE depends if the sample was cooled in or without a field. We note that 
the driving forces in a typical vibrating reed experiment are extremelly low --
shielding currents lower than $10^3$~A/m$^2$ -- which is an advantage in comparison
to other methods used to study pinning properties. Therefore, it could be
shown that at low enough vibrating amplitudes (low driving force) the PE
vanishes in a Y123 film when measured as a function of temperature at
constant field \cite{zieseprb}.  
A strong driving current dependence of the
PE has been also observed in 2H-NbSe$_2$ crystal \cite{hb}. In this crystal
the dissipation near the upper critical field strongly increases with the
transport current, in agreement with the observed behavior in \cite{zieseprb},
although the PE still remains \cite{hb}. 

Current driven order, disorder and depinning
of the FLL in 2H-NbSe$_2$ crystals have been studied in detail in 
Refs.~\cite{andrei1,andrei2,andrei3}. The influence of the probing current  on the 
first-order transition has been studied in Y123 crystals \cite{gor}.
In contrast to those studies, the technique 
used in this work allows
us to study the change of pinning using probing and driving 
currents substantially smaller than the critical 
one. The vibrating reed measurements provide
 the space derivative of the pinning force at small distances of the  
potential  minimum. 
 As we will show below, at low temperatures 
the measurements as a function of a driving dc current reveal an irreversible enhancement
of  the pinning  of the FLL. 
Loading a vibrating superconductor with a transport current  allows
us to resolve  very small changes in
the elastic coupling as a function of applied current far below the vortex
depinning transition $T < T_{D}(B_a)$ and applied forces much less than the critical ones.
 In this paper we show that a metastable state with a distinctly
higher level of pinning than that one obtained 
after field cooling procedure without a driving force can be reached  by 
applying a sufficiently large
driving current.
 Our work shows that the field-cooled state is not
a state of minimum potential energy for the FLL, at least for Y123 HTS, and therefore
  different metastable
 states may be obtained depending on the used experimental
procedure to study pinning-related properties.

\section{Experimental Details}
 The measuring setup
is schematically depicted in figure \ref{setup}. The 
samples (films with substrate) were glued onto a 
silicon single crystal host reed.
The dimensions of the host reed used in this study were  
(length $\times$ width $\times$ thickness) 
 $l \times w \times d = 5 \times 1 \times 0.1$ mm$^3$.
The reed was T-shaped and its back-side was covered with 15~nm
gold layer for the capacitive detection of the 
reed motion. The reed driving force supplied by a sinusoidal voltage 
was kept constant throughout
the measurements. Detailed description of the vibrating reed technique and 
the typical electronic arrangement are
published elsewhere \cite{Esquin91,ZEB94}. Our system allows us
to measure automatically the resonance frequency $\nu = \omega/2\pi$ of the reed
as a function of field, temperature or driving current density $J_T$.

The YBa$_2$Cu$_3$O$_7$ films were produced by pulsed laser 
deposition technique with a rotating substrate \cite{Lorenz}.
The films were deposited on a thin (0.1 mm) SrTiO$_3$ substrate to minimize
the total sample weight glued to the host reed. The results 
presented in this work were obtained with a film of dimensions:
 $d_p \simeq 400$ nm,  $w_p \simeq 1.2$ mm, and  
$l_p \simeq 1.6$ mm. 

The virgin field-cooled state was  achieved by
cooling down in a magnetic field through the superconducting transition temperature
to the temperature of the measurements without
appling a current load.
The maximum current load applied in the measurements was 100 mA. 
The measurements have been performed with the field applied parallel
to the CuO$_2$ planes of the film ($\theta = 0^{\circ} \pm 1^{\circ}$) and
at $\theta = 30^{\circ}$. No differences between the results obtained
for the two different angles were observed.

\section{Flux line tension and elastic pinning correction
in vibrating superconductors}
\begin{figure}
\centerline{\psfig{file=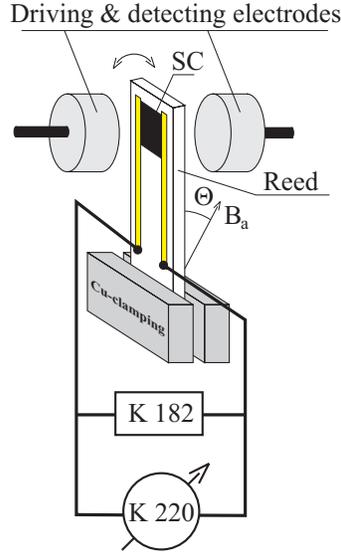,height=3.2in}}
\caption{Setup of the vibrating reed technique with the superconducting
film (SC). The film has two electrodes for electrical contacts used for injecting 
the transport current along the sample width and
perpendicular to the applied magnetic field direction $B_a$. The dc current
is supplied by a Keithley 220 source (K220). The voltage drop over the sample
is registred by a Keithley 182 voltmeter (K182).} 
\label{setup} \end{figure}

The resonance frequency enhancement of a vibrating superconductor
as a function of applied magnetic field $B_a$ is due to
the restoring force as a result of a pinned FLL. This behavior is well understood
and can be quantitatively accounted for by the theory developed in 
\cite{brandtvr}
(for a review see Refs. \cite{Esquin91,ZEB94}). 
We would like to outline here the main
results of the theory and restrict ourselves to the configuration
depicted in Fig. \ref{setup}
having the sample glued on a host reed and the magnetic field applied parallel
(or slightly tilted) to the length of the host reed and to the $ab$-planes of the films.

A superconducting sample with pinned FLL vibrating in a magnetic field 
distorts the field arround the superconductor and consequently  a positive
restoring force increases the resonance frequency of the cantilever.
In this case,  for a sample with dimension
$l_p \ge w_p >> d_p$ the magnetic line tension can be approximated
 by
\begin{equation}
P \simeq  \frac{\pi w_p}{4d_p} w_pd_p \frac{B_a^2}{\mu_0} \frac{l_p}{l} \,.
\label{p} \end{equation}
The condition of large enough pinning is satisfied for
\begin{equation}
l_p >> \Lambda_{44} \,,
\label{condit} \end{equation}
where 
$\Lambda_{44}=\lambda_{44}(\pi w_p/4 d_p)^{1/2}$
is the effective Campbell penetration depth for tilt waves and
$\lambda_{44} = (c_{44}/\alpha)^{1/2}$ with $c_{44} = B_a^2/\mu_0$  the FLL
tilt modulus. The Labusch parameter or elastic coupling
between FLL and the superconducting matrix is  
\begin{equation}
\alpha(B_a,T) = \frac{\partial^2}{\partial s^2}U(s,B_a,T)|_{s \rightarrow 0} = 
\left| \frac{F_P}{s_p} \right| \,,
\label{labusch}
\end{equation}
where $U$ is the mean value of the interaction energy per unit length (or 
pinning potential) between
vortices and pinning centers, $F_p = B_a J_c$ is the volume pinning force, 
$s$ is the flux line displacement relative to the superconducting matrix,  $s_p$
is the characteristic range of the pinning potential and $J_c$ the critical
current density.

The field dependence of the resonance frequency  depends on
the ratio of
the magnetic $W_M \simeq Pl$ to mechanical $W_R \simeq I\omega^2(0)$
 energies and the pinning strength of the FLL.
For  $X = 1.33(Pl/I\omega^2(0)) < 1$ and if (\ref{condit}) holds, the ideal enhancement
(infinite pinning case) of
the resonance frequency is given by \cite{ZEB94}
\begin{equation}
\omega_i^2(B_a) - \omega^2(0) \simeq Pl/I \,,
\label{Dw2} \end{equation}
where $I$
is the inertia moment of the cantilever with sample. This is  
 obtained experimentally at
low enough temperatures and applied fields where Eq.~(\ref{Dw2}) holds.
In the present work we determined
$I \simeq 5.8 \times 10^{-11}$ kg~m$^2$ for the used cantilever.

The field dependence of the resonance frequency is in general influenced
by the finite  pinning which is taken into account by a
correction term $\omega_{pin}^2$. Non-local  and
end effects due to the finite length of the sample are also taken
into account through the correction terms $\omega_{nl}^2$ and 
$\omega_{end}^2$, along with a correction term due to the finite
damping of the reed $\Gamma^2$. The field dependence of the 
resonance frequency can be then approximated by 
\cite{brandtvr}
\begin{equation}
\omega^2(B_a) - \omega^2(0) \simeq \frac{Pl}{I}-\omega_{pin}^2-\omega_{nl}^2-\omega_{end}^2-\Gamma^2 \,.
\label{corr} \end{equation}
In general, for the geometry of the used samples
 $\omega_{pin}^2 >> \omega_{nl}^2, \, \omega_{end}^2,
 \, \Gamma^2$. 

The correction term due to the finite pinning is given by \cite{brandtvr}
\begin{equation}
\omega_{pin}^2 \simeq 0.16  \omega^2(0) X \tau \,\, 
 \left (\frac{\Lambda_{44}}{l_p} \right )^2 
\left ( \frac{F(X')}{1+r/2} + \frac{pl_pG(X)}{w_p+p\Lambda_{44}\tau^{1/2}} \right) \,,
\label{wpin} \end{equation}
where $F(X)$ and $G(X)$ are tabulated numerical functions \cite{brandtvr},
$X'=X/(1+4 \beta X)$
with $\beta \simeq w_p^2/2\pi^2l_p^2$ ; $p \simeq 0.54 \ln(0.71w_p/d_p)$,
$r=(2X)^{1/2}\Lambda_{44}\tau^{1/2}/l_p$, and $\tau \simeq 1.08$ for
the case of not too strong pinning, when compressional waves penetrate
the superconductor entirely ($\Lambda_{44}>d_p$).

The deviation of the measured resonance frequency from the infinite
pinning value $\omega_i^2$ is inversely proportional to $\alpha$
via $\omega_{pin}^2 \propto \Lambda_{44}^2
\propto \alpha^{-1}$ and therefore it allows us to numerically calculate 
the Labush parameter
as a function of $B_a$. We should note, however, that Eq.~(\ref{wpin})
is only an approximation for the finite pinning correction and it does
not provide the correct value of  $\alpha$ for too large ($\Lambda < d_p$) or too weak
($\omega^2_{pin} \ge 0.1 Pl/I$) pinning.

In the temperature region of interest 90~K~$\ge T > 50~$K the Labusch
parameter increases approximately as $B_a^2$ having a value 
$\alpha(B_a = 1~{\rm T},~T = 73~{\rm K}) \simeq 4 \times 10^{17}$ N/m$^4$
in good agreement  with that obtained in Ref. \cite{ZEB94} for
YBa$_2$Cu$_3$O$_7$ films.

\section{Results and discussion}
\begin{figure}
\centerline{\psfig{file=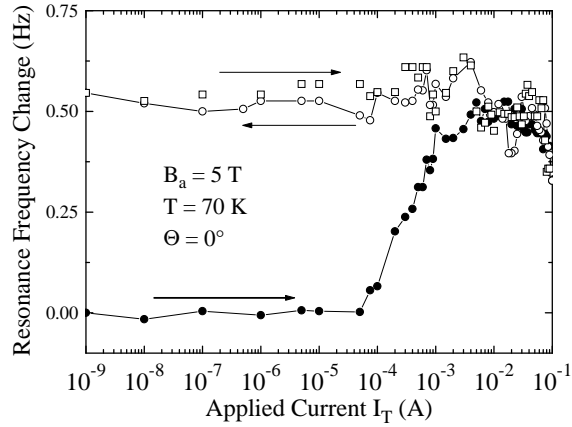,height=2.5in}}
\caption{The resonance frequency change as a function of the applied transport
current at $T = 70~$K, applied field $B_a = 5$ T (a) parallel to the sample
main surface ($\theta = 0^{\circ}$). Close symbols were obtained from the
virgin FC state increasing the current; the points with open symbols were obtained
by decreasing and increasing again the applied current.}
\label{irrever} \end{figure}
Figure \ref{irrever} shows the resonance frequency change
$\nu(I_T)-\nu(0)$ as a function of the applied driven current $I_T$
at constant field and temperature. The sample was 
cooled through the superconducting transition to 70~K in a field of 5~T 
applied parallel to the CuO$_2$ planes. The current
dependence of the virgin state is given by the close symbols in
Fig.~\ref{irrever}. 
We clearly observe that the resonance frequency
increases with applied current indicating, as explained in the last section, 
an increase of the elastic constant. The resonance frequency
reaches a maximum at a current $I_0 \sim 10^{-2}~$A which corresponds to
$\sim 10^{-4} J_c(70~$K,5~T). At larger currents the resonance frequency slightly decreases. 
Decreasing
the current, the resonance frequency follows  the virgin curve down to 
the current where the resonance frequency shows the maximum. For lower currents
the resonance frequency remains practically current independent.
This irreversible behavior is observed in the temperature and field range 
when the resonance frequency  increases with current
in the field-cooled 
virgin state. 

\begin{figure}
\centerline{\psfig{file=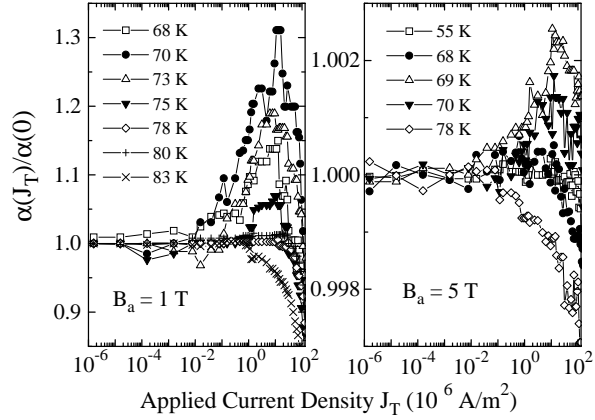,height=2.4in}}
\caption{Current dependence of the normalized Labusch parameter
measured at two different fields at different temperatures. Lines are only 
a guide to the eye.}
\label{25} \end{figure}
Figure \ref{25} shows the normalized elastic coupling as a function of 
the applied current density at two applied fields and at different constant
temperatures, calculated from the
measured resonance frequency change using Eqs.~(\ref{corr}) 
and  (\ref{wpin}). All curves were
measured starting from the field-cooled and zero-current virgin state.
Different behavior is observed at different temperatures of the measurement.
At low enough temperatures (e.g. $T < 60~$K) $\alpha$ remains
independent of the applied current up to high current densities (a slight decrease
of  $\alpha$ is observed at the highest currents). The current independence of
$\alpha$ at low $T$ is interpreted as due to the relatively large pinning which is 
not overwhelmed in the available applied current density range.

Increasing temperature (e.g. $T > 60~$K) , a maximum in $\alpha$ 
as a function of applied current clearly develops, see Fig.~\ref{25}. The 
increase in $\alpha$ reaches a maximum at a certain temperature (e.g., at $T \simeq 70~$K 
at $B_a = 1~$T). For higher temperatures, eventually $\alpha$ decreases with
applied current. In general, we can assume that the observed temperature 
dependence of $\alpha(I_T)$ results 
from the competing influence of
pinning and shear modulus $c_{66}(B_a,T)$ of the FLL at low temperatures and thermally
activated flux flow at high temperatures. We note that the absence of
a ``peak effect" in $\alpha$ as a function of current at high enough
temperatures does not imply the absence of the commonly observed
``peak effect" in $\alpha$ as a function of temperature at a constant
driving force. 

\begin{figure}
\centerline{\psfig{file=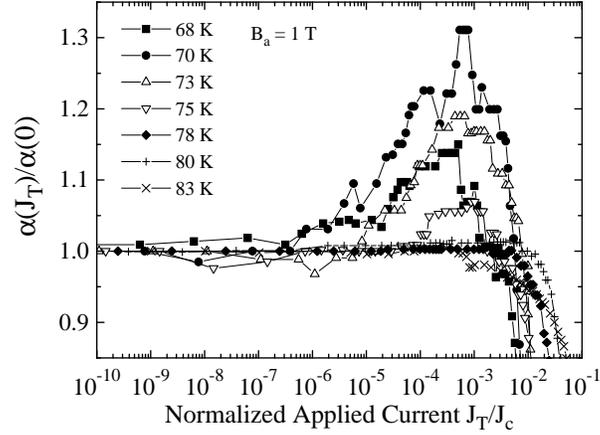,height=2.6in}}
\caption{Normalized Labusch parameter as a function of normalized
current density at $B_a = 1~$T and at different temperatures. 
$J_c$ is the critical current density.}
\label{1t} \end{figure}
Figure \ref{1t} shows the normalized elastic constant as a function
of the normalized applied current for a field of 1~T at different
temperatures. This behavior is similar to that observed for
other fields, see Fig.~\ref{25}. It is interesting to note that the
peak in $\alpha$ is reached at a current of the order of $10^{-3}$ of the
critical current density. The critical current density is obtained from
the $I-V$ characteristic curves and defined at the voltage $V = 2 \mu$V.
We note  that the rearrangement of 
the FLL is clearly observed already at currents several orders of 
magnitude smaller. We would like also to stress that the changes
in $\alpha$ are observed at a field (or temperature) much below
the corresponding depinning line $B_a \simeq 470[$T]$ (1 - t ) ^{1.4}$ with
$t = T_D(B_a)/T_D(0)$ measured at $I_T = 0~$, $T_D(0) = 88~$K \cite{pan}. The shift
of the depinning line with transport current is less than 2~K at the maximum
applied current of 100~mA and at fields $B_a \le 5~$T \cite{pan}.

The maximum increase in $\alpha$ with driving
current observed in our samples is of the order of $30 \%$, see
Fig.~\ref{1t}, in the measured field range ($B_a \ge 1~$T). This
increase agrees with the low-limit increase of pinning for defective 
FLL obtained by computer simulation studies \cite{brandtprl}. 
The FLL defects produced by the driving current are probably dislocations.
These are also created by thermal disorder and may lead to
an enhancement in the pinning strength when measured as a 
function of temperature.

The results clearly indicate that
the field cooled FLL is in a metastable, relatively ordered state. 
Upon field cooling the superconducting sample 
 through the superconducting and (thermally activated) depinning transitions,
the flux lines have to accomodate themselves in usually a random pinning
centers matrix  taking into account  the balance
between two contributions: vortex-vortex and 
vortex-pinning centers interactions. 
For our case of strong applied fields
$B_a >> B_{c1}$ we can assume that the
 vortex-vortex interaction overwhelms the pinning contribution
at temperatures above and  just below the depinning line. If  the driving
currents used to measure the pinning properties are small enough in order not to perturbate the FLL,
regions of the FLL remain
frozen in a metastable state. 
The observation of a peak effect
will depend therefore on the magnetic history and the perturbation
amplitude used to detect the pinning strength. The clear amplitude
- or driving force - and magnetic history dependence of the peak effect observed
near the depinning transition in Ref.~\cite{zieseprb} 
points out  the importance of defects in the FLL to produce the
peak effect. The driving current applied at constant field and 
temperature causes
the arrangement of the FLL into a more disordered state and the
elastic coupling increases. 
For intermediate fields, we expect that the higher the applied 
field the smaller will be the  increase of $\alpha$ produced by the driving current
since the vortex-vortex interaction increases whereas the current driven
production of the defects in the FLL weakens. This behavior is observed
experimentally, see Fig.~\ref{25}.
After reaching a maximum coupling
at a current $I_0$, the decrease of the resonance frequency can be
understood as due to the decrease of the current dependent pinning barrier.
This pinning barrier is field and temperature dependent and, therefore, the
observed decrease in $\alpha$ depends also on these parameters. 

Our results are in good agreement with the conclusions obtained by
transport and decoration experiments in 2H-NbSe$_2$ crystals \cite{dua}.
In this work \cite{dua} the authors found that the number of defects
in the vortex structure increases when the FLL starts to depin. The FLL becomes
nearly defect-free for large enough applied currents 
because the vortex-vortex interaction overwhelms. Our results show, see Fig.~4, 
that at large driven currents $J_T > 10^{-2}J_c$ the elastic coupling decreases
and therefore the pinning.
At high enough driven currents we expect a depinned and obviously more ordered 
FLL than any lattice in the FC state without driving forces.

At this point we would like to compare the published data on the peak
effect measured with the vibrating reed technique in Y123 films and
crystals \cite{zieseprb} as well as the peak effect and current driven (re)organization
of the FLL observed in 2H-NbSe$_2$ crystals \cite{andrei1,andrei2,andrei3}.
Vibrating-reed results in Y123 thin films showed that the enhancement of $\alpha$
as a function of temperature at constant applied field and in the field-cooled (FC) state 
is observed only if the driving force (or vibrating amplitude of the 
superconducting sample) is large enough. In contrast to the FC state, the
zero-field cooled (ZFC) state shows  a weak or no signature of a pinning enhancement near
the depinning line.

For the case the peak effect in $\alpha$ is observed near the depinning
temperature $T_D(B_a)$, the results indicate that $\alpha$ in the ZFC state 
and near $T_D(B_a)$ is smaller than $\alpha$ in the FC state \cite{zieseprb}. 
This observation agrees qualitatively
with the smaller critical current density measured in the ZFC 
state in 2H-NbSe$_2$ at temperatures below
the peak effect \cite{andrei2,andrei3}.
From this observation, the authors in Refs.~\cite{andrei2,andrei3} proposed
that the ZFC state has a more ordered FLL than the FC state, arguing that in
the $I_c$ measurements in the ZFC state the vortices enter the sample with large
velocities which lead to a more ordered FLL state. 
We note, however, that in the case of the vibrating reed, the perturbation to the FLL is kept
much smaller than that used for $I_c$ measurements. 
Therefore, we think that, in
general, the ZFC state has a less homogeneous field distribution in the sample than in
the FC state. Increasing temperature, an effective smaller $\alpha(B_a,T)$ is measured
in the ZFC state because: (a)  the effective field in the sample is smaller and (b) there is a
change of the effective pinning potential due to the field gradient. The differences
between the ZFC and FC states vanish as soon as the thermal creation of FLL defects and
the flux creep compensate the inhomogeneous field distribution.

\section{Conclusion}
We have shown that the field-cooled state in 
YBa$_2$Cu$_3$O$_7$ superconducting films can be driven to
a lower potential equilibrium state at fields and temperatures far below 
the depinning transition. This new equilibrium state of the FLL has
a larger elastic coupling.  Our results indicate that after field cooling without 
driving forces, the static FLL is in a metastable, relatively {\em ordered} state. 
Its pinning can be enhanced by the
current driven creation of defects in the FLL. Our results support the recently
published interpretation on the ``melting" transition that argues that the FLL does not melt at the
depinning line but the opposite, it decouples from the superconducting matrix
becoming more ordered \cite{mel}. This ordering is partially retained in the
FLL after FC through the depinning line without driving forces.

Acknowledgements: 
We would like to acknowledge M. Lorenz 
for providing us with the YBa$_2$Cu$_3$O$_7$
superconducting films and Y. Kopelevich for a careful reading
of the manuscript and fruitful discussions.
This work is supported by the Deutsche Forschungsgemeinschaft within 
the ``Innovationskolleg Ph\"anomene an der Miniaturisierungsgrenze" 
(Project H, DFG IK 24/B1-1) and the German-Israeli Foundation for Scientific Research and Development 
(Grant G-0553-191.14/97).

\end{document}